\documentclass[a4paper,fleqn,usenatbib]{mnras}

\usepackage{txfonts}
\usepackage[T1]{fontenc}
\usepackage{ae,aecompl}

\usepackage{graphicx}	

\title[Chemical composition of NGC\,4609 and NGC\,5316]{Chemical composition of evolved stars in the young open clusters NGC\,4609 and NGC\,5316 \thanks{Based on observations collected at the European Organisation for Astronomical Research in the Southern Hemisphere under ESO programme 085.D-0093(A).}}

\author[A. Drazdauskas et al.]{
Arnas Drazdauskas,$^{1}$\thanks{E-mail: arnas.drazdauskas@tfai.vu.lt}
Gra\v{z}ina Tautvai\v{s}ien\.{e},$^{1}$
Rodolfo Smiljanic,$^{2}$
\newauthor Vilius Bagdonas,$^{1}$
 and Yuriy Chorniy$^{1}$
\\
$^{1}$Institute of Theoretical Physics and Astronomy, Vilnius University, Saul\.{e}tekio al. 10222 Vilnius, Lithuania\\
$^{2}$Nicolaus Copernicus Astronomical Center, Polish Academy of Sciences, Bartycka 18, 00-716, Warsaw, Poland\\
}

\date{Accepted 2016 July 12. Received 2016 July 1}

\pubyear{2016}

\begin{document}
\label{firstpage}
\pagerange{\pageref{firstpage}--\pageref{lastpage}}
\maketitle

\begin{abstract}
High-resolution spectral analysis is performed for the first time in evolved stars of two young open clusters: 
NGC\,4609 and NGC\,5316, of about 80 and 100 Myr in age, respectively, and turn-off masses above 5~$M_{\odot}$. Stellar evolution models predict an extra-mixing event in evolved stars, which follows the first dredge-up and  happens later on the red giant branch. However, it is still not understood how this process affects stars of different masses. In this study, we determine abundances of the mixing sensitive elements carbon and nitrogen, carbon isotope  $^{12}{\rm C}/^{13}{\rm C}$ ratios, as well as 20 other elements produced by different nucleosynthetic processes (O, Na, Mg, Al, Si, Ca, Sc, Ti, Cr, Mn, Co, Ni, Y, Zr, Ba, La, Ce, Pr, Nd, and Eu). We compared our results with the latest theoretical models of evolutionary mixing processes. We find that the obtained $^{12}{\rm C}/^{13}{\rm C}$ and C/N ratios and [Na/Fe] agree quite well with the model which takes into account thermohaline- and rotation-induced mixing but within error limits also agree with the standard first dredge-up model. Comparison of oxygen, magnesium and other $\alpha$-elements with theoretical models of Galactic chemical evolution revealed that both clusters follow the thin disk $\alpha$-element trends. Neutron-capture element abundances in NGC\,4609 are apparently reflecting its birthplace in the thin disk, while NGC\,5316 has marginally higher abundances, which would indicate its birthplace in an environment more enriched with neutron-capture elements.
\end{abstract}




\section{Introduction}

We continue investigations of evolutionary mixing processes in evolved stars of open clusters (OCs; \citealt{Mikolaitis12, Tautvaisiene15, Drazdauskas16} and references therein). The study of stars in OCs gives us many advantages when compared to a single object analysis. With a sample of stars of roughly the same origin (age, distance, metallicity), we can greatly improve the accuracy of the analysis. OCs span a variety of ages, distances and positions in the Galaxy. They have stars in different stages of evolution, which is beneficial when trying to understand evolutionary effects on the changes of the photospheric chemical composition.

Analysing the chemical composition of stars can provide not only insights into stellar evolution, but into the evolution of the Galaxy as well. OCs with multiple objects can provide us with better determinations of their positions in the Galaxy (distance from the Sun and the centre of the Galaxy) and their ages than we can derive for single stars. Having multiple stars in a single OC can provide better statistics for determination of chemical abundances. Combining the more precise positions, with accurate chemical element abundance determinations gives us an opportunity to study abundance gradients in our Galaxy. Among the most important galactic evolution indicators are the $\alpha$-elements. According to numerous studies, there is a significant difference between $\alpha$-element abundances in different parts of the Galaxy (\citealt{Bensby14, Neves09} and references therein). In addition to light chemical elements, we determine heavy element abundances as well. The so called s- and r-process elements are produced during neutron capture processes (\citealt{Burbidge57}) at different stages of stellar evolution. It is well known that asymptotic giant branch (AGB) stars are mainly responsible for synthesis of s-process elements. However, their influence on production of these elements was underestimated, especially in young clusters \citep{D'Orazi09, Jacobson13}. The site of the r-process nucleosynthesis is still under debate, possibilities include core collapse supernovae (e.g., \citealt{Woosley94, Nishimura15}) and neutron star mergers \citep{Eichler89, Freiburghaus99}. Comparing the neutron capture element abundances with the metallicity, we can gain valuable insights into the evolution of our Galaxy (\citealt{Maiorca11, Maiorca12, Reddy12, Reddy13, Reddy15, Mishenina15, Overbeek16}).

\begin{figure}
	\includegraphics[width=\columnwidth]{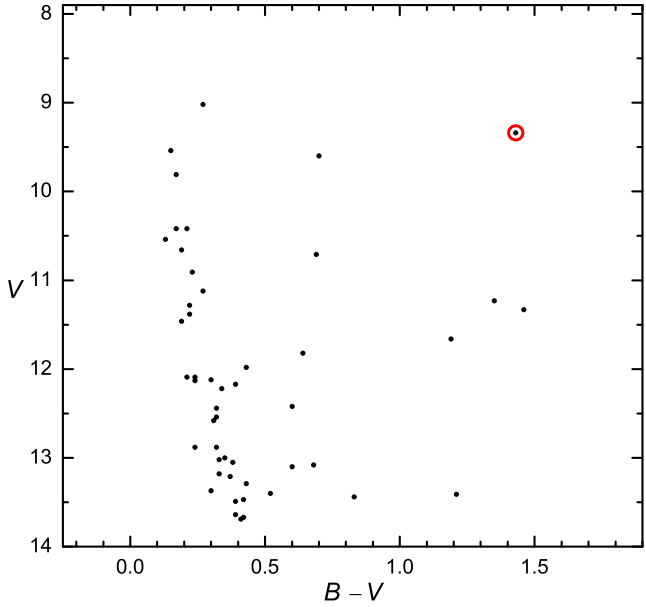}
    \caption{The colour-magnitude diagram of the open cluster NGC\,4609. The star investigated in this work is indicated by the open circle. The diagram is based on $UBV$ photometry by \citet{Feinstein71}}
    \label{fig:1}
\end{figure}

\begin{figure}
	\includegraphics[width=\columnwidth]{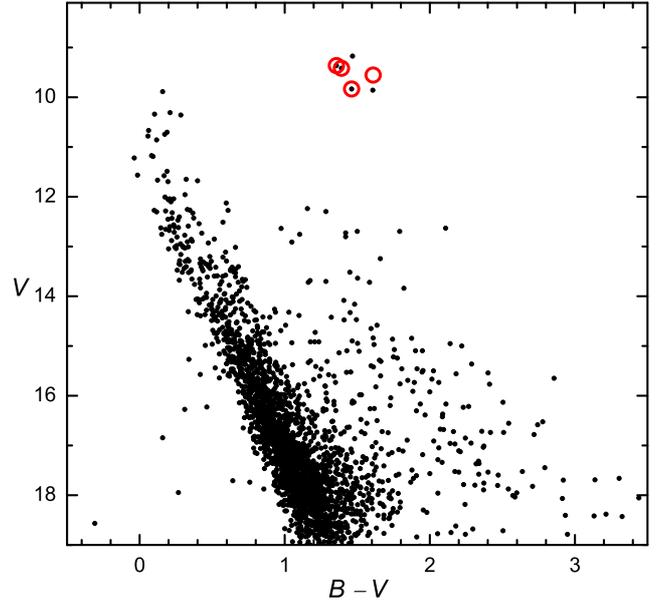}
    \caption{The colour-magnitude diagram of the open cluster NGC\,5316. The stars investigated in this work are indicated by open circles. The diagram and the three stars are based on $UBV$~CCD photometry by \citet {Carraro12} and the star marked by empty circle is from T.~Oja (private communication)}
    \label{fig:2}
\end{figure}

Our main goal is to determine abundances of mixing sensitive chemical elements, and carbon and nitrogen in particular. Through CNO cycle, CNO elements play an important role as energy sources in stellar interiors. This affects stellar positions in a Hertzsprung-Russel (HR) diagram and the production of heavy chemical elements. The carbon and nitrogen abundances, C/N, and carbon isotope ratios $^{12}{\rm C}/^{13}{\rm C}$ are important tools to study stellar evolution. When low- and intermediate-mass stars are at the bottom of the giant branch, the first dredge-up takes place \citep{Iben65}. Surface abundances are further modified when extra-mixing processes become efficient. This happens when low-mass stars reach the luminosity bump on the red giant branch (RGB). Variations of $^{12}{\rm C}/^{13}{\rm C}$ and C/N ratios depend on several factors: stellar evolutionary stage, mass, and metallicity \citep{Gilroy89, Luck94, Gratton00, Tautvaisiene00, Tautvaisiene05, Tautvaisiene10, Smiljanic09, Smiljanic16, Mikolaitis10, Mikolaitis11a, Mikolaitis11b, Mikolaitis12}. On the other hand, exact dependencies are still uncertain \citep{Chaname05, Charbonnel06, Cantiello10, Charbonnel10, Denissenkov10, Lagarde11, Lagarde12, Wachlin11, Angelou12, Lattanzio15}. There are models which predict different outcomes of mixing effects depending on the stellar turn-off (TO) mass. In this work, we try to fill a lack of observational data in the TO mass range >5~$M_{\odot}$ by analysing two relatively young open clusters.

The first-targeted cluster in our work is NGC\,4609. There are only a few previous photometric studies and no spectroscopic ones for this cluster. The first ever study of NGC\,4609 was by \citet{Feinstein71}. Their photoelectric $UBV$ observations and analysis of the cluster provided an age of 60~Myr and a distance of 1320~pc as well as identified 33 true members. Consecutive age determinations range from 36 Myr (\citealt{Battinelli91}) to 125 Myr (\citealt{Kharchenko13}). The following photometric study by \citet{Mermilliod81} confirmed that the distance is about 1325~pc,  and provided values for the reddening $E(B-V) = 0.34$ and extinction $(m-M) = 11.68$. \citet{Jura87} made a rough estimate of TO mass to be around 7~$M_{\odot}$. The only metallicity determination was made by \citet{Claria89}. They obtained $UBV$, David Dunlap Observatory($\rm DDO$) and $\rm Washington$ photometry for G and K stars and found a metallicity of about $0.05 \pm 0.13$.

The second open cluster analysed in this work is NGC\,5316. The first photometric age and distance determination was made by \citet{Lindoff68}. They derived an age of 51~Myr and a distance around 1.12~kpc. The following photometric analysis of the cluster provided similar distance values, ranging from 1.08~kpc (\citealt{Becker71}) to 2~kpc (\citealt{Chen03}). The determined Galactocentric distances do not vary much as well (from 7.4~kpc by \citealt{Chen03} to 7.84~kpc by \citealt{Strobel91}). Age determinations however vary more significantly from the already mentioned determination by \citet{Lindoff68} to around 195~Myr determined by \citet{Battinelli91}. The most recent determination of about 170~Myr comes from \citet{Kharchenko13}. The newest most comprehensive study by \citet{Carraro12} from $UBVI$ CCD photometry provided an age of 100~Myr, a distance of 1.4~kpc and $R_{\rm gc}$ of 7.6~kpc, as well as values for reddening $E(B-V) = 0.25 \pm 0.05$ and extinction $(m-M) = 11.50 \pm 0.2$. The only TO mass determination was made by \citet{Jura87}, who provides a value of about 4~$M_{\odot}$.

NGC\,5316 has numerous photometric metallicity determinations. The first one being by \citet{Claria89} who used multiple photometric systems to derive $\rm [Fe/H] = 0.19 \pm 0.11$ from seven RGB stars. The following studies by \citet{Piatti95} and \citet{Twarog97} used the same $DDO$ photometry as \citet{Claria89}, however different analysis methods were applied. The resulting derived abundances were $\rm [Fe/H] = -0.02 \pm 0.12$ and $\rm [Fe/H] = 0.13 \pm 0.13$, respectively. The newest study by \citet{Kharchenko13} provides the metallicity of $\rm [Fe/H] = 0.05 \pm 0.13$ based on two stars.

\section{Observations}

\begin{table*}
	\centering
	\caption{Parameters of the programme stars and a log of observations}
	\label{table:1}
	\begin{tabular}{lcccccccccc}
		\hline\hline
		\noalign{\smallskip}
ID & $V$ & $B-V$ & RV$_{\rm This\,work}$& $\sigma$ & RV$_{\rm WEBDA}$ & R.A. & DEC & Time obs. & Exp. time & S/N \\
   & mag & mag   &km s$^{-1}$& km s$^{-1}$&km s$^{-1}$&deg (J2000)   & deg (J2000)    &           &     seconds      &     \\
  		\hline
		\noalign{\smallskip}
	\multicolumn{11}{|c|}{NGC\,4609}\\
43 & 9.34 & 1.43 & $-20.74$ & 0.56 & $-21.30$ & 301.9865   &  $-00.0553$   & 2010-06-23 & 600 & $\sim$180 \\
	\noalign{\smallskip}
	\multicolumn{11}{|c|}{NGC\,5316}\\	
31 & 9.39 & 1.43 & $-13.50$ & 0.75& $-14.71$ &310.1919 & +00.1544 & 2010-06-25 & 600 & $\sim$200 \\
35 & 9.44 & 1.48 & $-15.03$ & 0.60 & $-15.53$ &310.1819 & +00.1307 & 2010-06-25 & 600 & $\sim$200 \\
45 & 9.55 & 1.61 & $-13.79$ & 0.66 & $-14.45$ &310.2426 & +00.1333 & 2010-06-27 & 600 & $\sim$200 \\
72 & 9.84 & 1.59 & $-15.52$ & 0.70 & $-15.72$ &310.2247 & +00.1239 & 2010-06-27 & 900 & $\sim$150 \\
\hline		
	\end{tabular}
\end{table*}

The observations were carried out using bench-mounted, high-resolution, astronomical \'{E}chelle spectrograph FEROS (Fiber-fed Extended Range Optical Spectrograph; \citealt{Kaufer99}) at the 2.2~m Max Planck Gesellschaft/European Southern Observatory (ESO) Telescope in La Silla, between 2010 June 25 and 27. FEROS provides a full wavelength coverage of 3500--9200~\AA\ over 39 orders with a resolving power of about 48000. All spectra were reduced with the FEROS DRS (Data Reduction System) pipeline within MIDAS. 

A total of 5 stars were observed (one in NGC\,4609 and four in NGC\,5316). As can be seen in Figs.~\ref{fig:1} and~\ref{fig:2}, the chosen stars are at the position of the red clump. We present parameters of the observed stars and a log of observations in Table~\ref{table:1}. The magnitudes were taken from the same sources as for the CMD diagrams (see Figs.~\ref{fig:1} and~\ref{fig:2}). The heliocentric radial velocities were computed in this work using DAOSPEC \citep{Stetson08} with the Gaia-ESO line list \citep{Heiter15} and the IRAF task RVCORRECT. Radial velocity values from the WEBDA database \citet{Mermilliod08} are listed as well and are in agreement with our results.

\begin{figure}
	\includegraphics[width=\columnwidth]{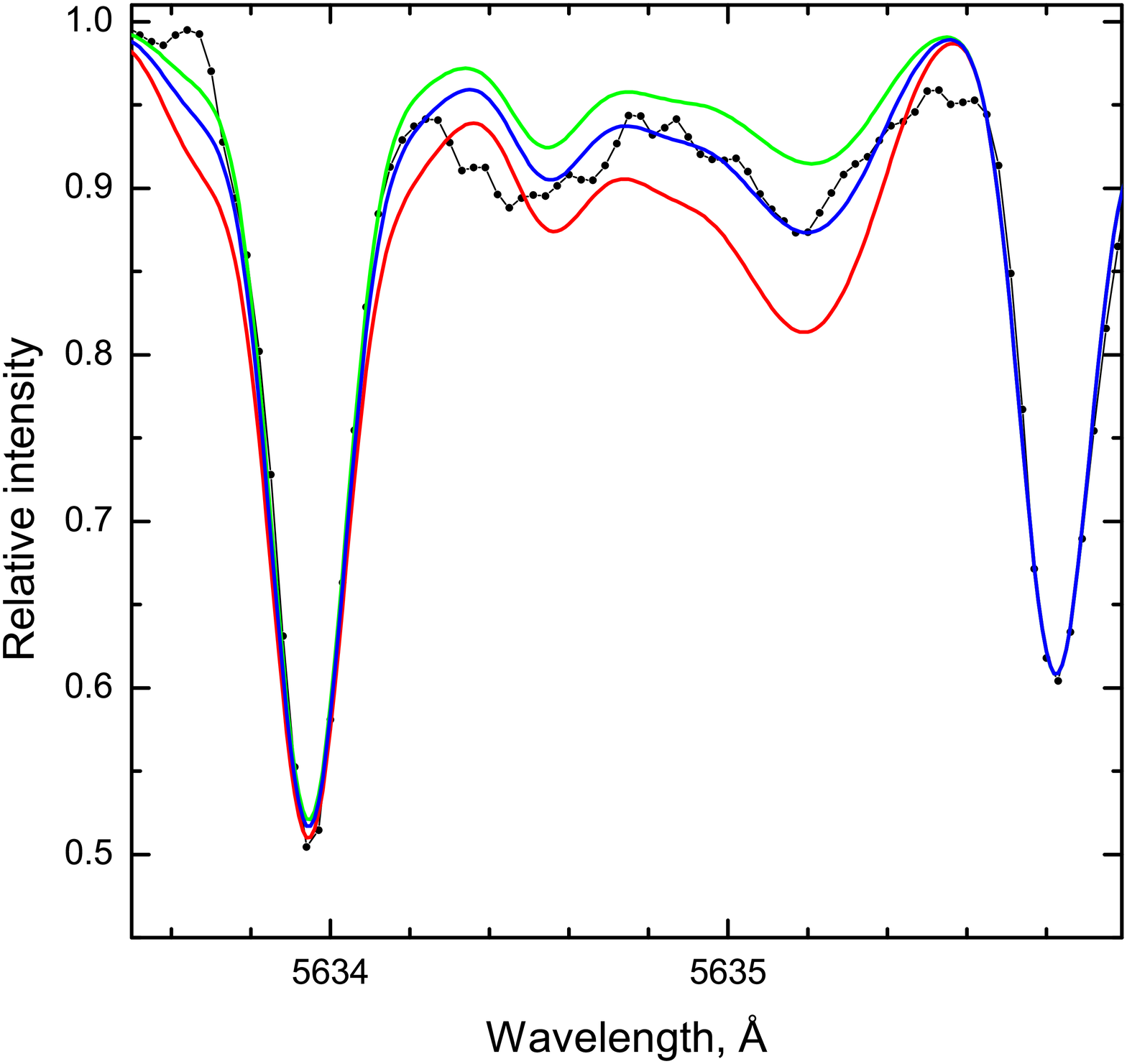}
    \caption{A fit to the ${\rm C}_2$ Swan (0,1) band head at 5635.5~{\AA} in the star NGC\,5316\,45.The observed spectrum is shown as a black line with dots. The synthetic spectra with ${\rm [C/Fe]}=-0.08\pm 0.1$ are shown as solid lines.}
    \label{fig:3}
\end{figure}

\begin{figure}
	\includegraphics[width=\columnwidth]{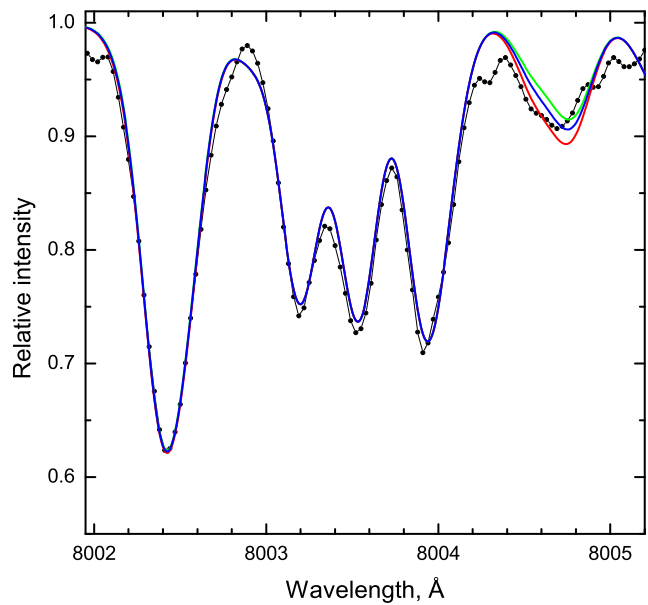}
    \caption{A fit to the CN bands in the star NGC\,4609\,43. The observed spectrum is shown as a black line with dots. The synthetic spectra with $^{12}{\rm C}/^{13}{\rm C}$ ratio $23 \pm 5$ are shown as solid lines.}
    \label{fig:4}
\end{figure}

\subsection{Membership}

NGC 4609 is a poorly populated cluster with only one red giant in its field (NGC\,4609\,43 or HD\,110478). A comprehensive radial velocity membership study has never been performed for this cluster. The radial velocity monitoring of the star 43 by \citet{Mermilliod08} suggests that this is a single star. The proper motion analysis of many open clusters by \citet{Dias14} includes stars in NGC\,4609, but does not include the star 43. Only photometric membership analyses have been performed. The first photometric study of this cluster remains as its most comprehensive photometric study \citep{Feinstein71}. In that paper, photometry was obtained for 52 stars out of which 33 are considered to be members (including the red giant), although the membership criteria are not explained. \citet{Claria89A} further obtained and analyzed $UBV$, $\rm DDO$, and $\rm Washington$ photometry of this giant and concluded that it is a cluster member based on two criteria. The first compares the reddening computed for the red giant with that obtained for the main-sequence stars. The second compares the predicted luminosity class of the giant, given the cluster distance, with that derived using $\rm DDO$ indices. According to \citet{Claria89}, the results of both methods indicate that the star 43 is a member of NGC\,4609.

The photometric studies by \citet{Rahim66} and \citet{Carraro12}, both down to $V \sim 16$, find up to seven candidate red giants in the field of NGC\,5316. The four stars from NGC\,5316 that are analyzed in this work were originally chosen from the radial velocity study of \citet{Mermilliod08}. These authors monitored six red giants in the field of the cluster. The four stars (NGC\,5316\,31, 35, 45, and 72) were found to be members, while the remaining two are non-members and spectroscopic binaries. The analysis of Tycho\,2 proper motions by \citet{Dias02} found that the stars NGC\,31, 35, 45, and 72 have membership probabilities of 83\%, 76\%, 78\%, and 86\%, respectively. Moreover, we note that \citet{Claria89B} analyzed $UBV$, $\rm DDO$, and $\rm Washington$ photometry of seven giants in the field of NGC\,5316. These authors applied the same two membership methods discussed above for NGC\,4609, and found that only the four stars analyzed in our work are confirmed members of NGC\,5316. 

\section{Method of analysis}
\label{sec:3}

Atmospheric parameters of stars were determined using the standard spectroscopic method. 
The effective temperature ($T_{\rm eff}$) was determined requiring that the abundance of Fe\,{\sc i} lines did not depend on the lower level excitation potential ($\chi$). The surface gravities (log~$g$) were determined from the iron ionization equilibrium. The microturbulence velocity ($v_{\rm t}$) was found by minimising a scatter between abundances of Fe\,{\sc i} lines and assuming that there should be no relation between abundances and equivalent widths. Determining these three main parameters through multiple iterations provides us with the value of metallicity as calculated from the used Fe\,{\sc i} lines. We used between 22 and 29 Fe\,{\sc i} lines to derive the atmospheric parameters and metallicity content for the target stars.

We analysed the spectra using a differential analysis technique. All calculations were differential with respect to the Sun. Solar element abundance values were taken from \citet{Grevesse02}. For the abundance calculations which were based on equivalent widths of spectral lines we used the EQWIDTH code, and for those which were based on spectral synthesis we used the BSYN code, both developed at the Uppsala Observatory.  A set of plane-parallel, one-dimensional, hydrostatic, constant flux LTE model atmospheres were taken from the MARCS stellar model atmosphere and flux library\footnote{http://marcs.astro.uu.se/} described by \citet{Gustafsson08}.

We used equivalent widths for the determination of abundances for Na, Mg, Al, Si, Ca, Sc, Ti, Cr, Mn, Co, and Ni chemical elements. Each line for these elements was inspected and some of them were excluded due to blending or other contamination effects, hence the line list differs slightly for every star. For this task we used the SPLAT-VO \citep{Skoda14} tool developed by the German Astrophysical Virtual Observatory and the Astronomical Institute of the Academy of Sciences of the Czech Republic.  
For abundance calculations for the mentioned elements we used neutral lines, however we determined the Ti\,{\sc ii} abundance from ionized lines as well. 

Spectral synthesis was used for C, N, O, and  neutron-capture chemical element abundance determinations as well as for $^{12}{\rm C}/^{13}{\rm C}$\ ratio calculations.  The Vienna Atomic Line Data Base (VALD, \citealt{Piskunov95}) was used to prepare input data for the calculations and the atomic oscillator strengths for the lines used were taken from the inverse solar spectrum analysis by \citet{Gurtovenko89}.

For the carbon abundance determination in all stars we used two regions: ${\rm C}_2$ Swan (0,1) band heads at 5135.5~{\AA} and 5635.2~{\AA}. We used the same molecular data of ${\rm C}_2$  as by \citet {Gonzalez98}. The oxygen abundance was derived from synthesis of the forbidden [O\,{\sc i}] line at 6300~{\AA}. The $gf$\ values for $^{58}{\rm Ni}$ and $^{60}{\rm Ni}$ isotopic line components, which blend the oxygen line, were taken from \citet{Johansson03}. An interval of 7980-8010~{\AA} containing strong 
CN features was used to determine nitrogen abundance and $^{12}{\rm C}/^{13}{\rm C}$\ ratios. The $^{12}{\rm C}/^{13}{\rm C}$\ ratio was obtained from the $^{13}{\rm C}/^{12}{\rm N}$ feature at 8004.7~{\AA}. The CN molecular data for this wavelength interval were provided by Bertrand Plez. Figs.~\ref{fig:3} and \ref{fig:4} display examples of spectrum syntheses for the programme stars. The best-fitting abundances were determined by eye.

The sodium abundance was determined from three or four lines (4751.8, 5682.6, 6154.2 and 6160.8~\AA). We applied non-local thermodynamic equilibrium (NLTE) corrections as described in the paper by \citet{Lind11}. The corrections range from $-0.09$~dex in the star NGC\,5316\,35 to $-0.12$~dex in NGC\,5316\,31 and NGC\,5316\,72. The magnesium abundance was  determined from equivalent widths of two or three Mg\,{\sc i} lines (5711.09, 6318.70, and 6319.24~\AA). 

Abundances for neutron-capture chemical elements in our programme stars were derived using synthetic spectra. We used ionized lines for elements Y, Ba, La, Ce, Pr, Nd, Eu and neutral - for Zr. Y was determined from the Y\,{\sc ii} lines at 4883.69, 4900.12, 4982.14, 5200.41, 5402.78~\AA; Zr from the Zr\,{\sc i} lines at 5385.1 and 6127.5~\AA; Ba from the Ba\,{\sc ii} lines at 6141.00 and 6496.91~\AA; La from the La\,{\sc ii} lines at 4748.73, 6320.41, 6369.48 \AA; Ce from the Ce\,{\sc ii} lines at 4773.96, 5274.22, 5512.00, 5610.3, 6043.00~\AA; and Nd abundance was derived from the Nd\,{\sc ii} lines at 5092.8, 5255.52, 5293.2, 5319.8~\AA. After careful line inspection, we have chosen between one and five lines for each element. Every spectral line fit was inspected individually and after such inspection some of the lines were excluded due to their apparent asymmetry or particularly high deviation of abundance from remaining lines for that element. 

A hyperfine structure (HFS) was taken into account for the synthesis of barium line at 6496.9~\AA\ \citep{McWilliam98} as well as for the Pr\,{\sc ii} lines  at 5259.7 and 5322.8~\AA \citep{Sneden09}. The europium abundance was determined from a single Eu\,{\sc ii} feature at 6645.1~\AA. The wavelength, excitation energy and total log~$gf = 0.12$ were taken from \citet{Lawler01}, the isotopic fractions of $^{151}{\rm Eu}$ 47.77\% and $^{153}{\rm Eu}$ 52.23\%, and isotopic shifts were taken from \citet{Biehl76}. We also took into account the partial blending of the said line with Si\,{\sc i} and Ca\,{\sc i} lines at 6645.21 \AA.

Since stellar rotation can be important when analysing young stars, we estimated it by looking at stronger surrounding lines in spectral regions around the investigated C, N, and O features. Our roughly estimated values are presented together with the atmospheric parameters in Sect.~\ref{sec:Results}.

\begin{table}
	\centering
    \caption{Effects on derived abundances, $\Delta$[A/H], resulting from model changes for the star NGC\,4609\,43. }
    \label{table:2}
    \begin{tabular}{lrrcccc}
    \hline
    \noalign{\smallskip}
Species & ${ \Delta T_{\rm eff} }\atop{ \pm100~{\rm~K} }$ & ${ \Delta \log g }\atop{ \pm0.3 }$ & ${ \Delta v_{\rm t} }\atop{ \pm0.3~{\rm km~s}^{-1}}$ &
             ${ \Delta {\rm [Fe/H]} }\atop{ \pm0.1}$ & Total \\
    \noalign{\smallskip}
    \hline
    \noalign{\smallskip}
C (C$_{2}$) &			0.01 	& 	0.08 	& 	0.00 	&	0.02 	&	0.08    \\
N (CN) &			0.05 	& 	0.07 	& 	0.01 	&	0.02 	&	0.09    \\
O ({[O\,{\sc i}]}) &			0.01 	& 	0.13 	&	0.00 	&	0.03 	&	0.13    \\
$^{12}{\rm C}/^{13}{\rm C}$ & 	1	& 	1 	& 	0 	&	0 	&	1.4   	\\
Na\,{\sc i} (LTE)	&	$0.09$	&	$0.04$	&	$0.10$	&	$0.03$	&	0.14\\
Mg\,{\sc i}	&	$0.02$	&	$0.02$	&	$0.04$	&	$0.01$	&	0.05	\\
Al\,{\sc i}	&	$0.08$	&	$0.01$	&	$0.07$	&	$0.02$	&	0.10	\\
Si\,{\sc i}	&	$0.05$	&	$0.07$	&	$0.08$	&	$0.01$	&	0.11\\
Ca\,{\sc i}	&	$0.11$	&	$0.01$	&	$0.13$	&	$0.02$	&	0.16	\\
Sc\,{\sc i}	&	$0.02$	&	$0.13$	&	$0.11$	&	$0.04$	&	0.17	\\
Ti\,{\sc i}	&	$0.14$	&	$0.01$	&	$0.13$	&	$0.02$	&	0.19	\\
Ti\,{\sc ii}&	$0.02$	&	$0.13$	&	$0.13$	&	$0.04$	&	0.19	\\
Cr\,{\sc i}	&	$0.10$	&	$0.01$	&	$0.15$	&	$0.02$	&	0.18	\\
Mn\,{\sc i}	&	$0.09$	&	$0.06$	&	$0.16$	&	$0.03$	&	0.19	\\
Co\,{\sc i}	&	$0.04$	&	$0.06$	&	$0.15$	&	$0.02$	&	0.17\\
Ni\,{\sc i}	&	$0.01$	&	$0.06$	&	$0.16$	&	$0.11$	&	0.17	\\
Y\,{\sc ii}	&	$0.01$	&	$0.13$	&	$0.07$	&	$0.04$	&	0.15\\
Zr\,{\sc i}	&	$0.20$	&	$0.02$	&	$0.06$	&	$0.01$	&	0.21\\
Ba\,{\sc ii}	&	$0.05$	&	$0.14$	&	$0.45$	&	$0.04$	&	0.47\\
La\,{\sc ii}	&	$0.02$	&	$0.14$	&	$0.01$	&	$0.03$	&	0.14\\
Ce\,{\sc ii}	&	$0.02$	&	$0.13$	&	$0.08$	&	$0.05$	&	0.15\\
Pr\,{\sc ii}	&	$0.03$	&	$0.13$	&	$0.00$	&	$0.05$	&	0.14\\
Nd\,{\sc ii}	&	$0.03$	&	$0.13$	&	$0.19$	&	$0.04$	&	0.23\\
Eu\,{\sc ii}	&	$0.01$	&	$0.14$	&	$0.01$	&	$0.04$	&	0.14\\
    \noalign{\smallskip}
    \hline
    \end{tabular}
\end{table}

Two types of uncertainties should be taken into account. First, we have the errors that affect each line independently: uncertainties of atomic parameters, placement of the local continuum, and the fitting of the line itself. Secondly, there are errors that affect all lines at the same time (such as uncertainties in the stellar atmospheric parameters and the model used). 

Table~\ref{table:2} shows a relation between the abundance estimates [El/Fe] and assumed uncertainties of the atmospheric parameters in the programme star NGC\,4609\,43. Considering the given deviations from the used parameters, we see that the abundances of the majority of chemical elements are not affected strongly. However, Ba\,{\sc ii} lines are very sensitive to the microturbulence velocity for a set of atmospheric parameters of this star.

A scatter of the deduced line abundances gives an estimate of the uncertainty due to the random errors. The uncertainties in the derived abundances that are the result of random errors amount to approximately $0.06$~dex.

Since abundances of C, N, and O are also bound together by the molecular equilibrium in the stellar atmospheres, we further investigated how an error in one of them typically affects the abundance determination of another. Thus $\Delta{\rm [O/H]}=0.10$ causes $\Delta{\rm [C/H]}=0.04$ and $\Delta{\rm [N/H]}=0.08$; $\Delta{\rm [C/H]}=0.10$ causes $\Delta{\rm [N/H]}=-0.10$ and $\Delta{\rm [O/H]}=0.02$; $\Delta {\rm [N/H]}=0.10$ has no effect on either the carbon or the oxygen abundances.

\section{Results and discussion}
\label{sec:Results}
\subsection{Parameters of open clusters}

\begin{table*}
	\centering
	\caption{Main atmospheric parameters of programme stars}
	\label{table:3}
	\begin{tabular}{lccccccccc}
		\hline\hline
		\noalign{\smallskip}
ID & $T_{\rm eff}$ & log~$g$ & $v_{\rm t}$ & $\upsilon~ \rm sin~ \textit{i}$ &[Fe/H] & $\sigma_{\rm FeI}$ & $n$ & $\sigma_{\rm FeII}$ & $n$ \\
  & K && km s$^{-1}$ &km s$^{-1}$&&\\
  		\hline
		\noalign{\smallskip}
	\multicolumn{9}{|c|}{NGC\,4609}\\
43 & 4620 & 2.20 & 1.45 & 2.0 &0.16 & 0.08 & 29 & 0.03 & 6 \\
	\noalign{\smallskip}
	\multicolumn{9}{|c|}{NGC\,5316}\\	
31 & 4825 & 1.70 & 1.85 & 6.9 &$0.05$ & 0.05 & 28 & 0.04	& 4	\\
35 & 4650 & 2.00 & 2.05 & 5.2 &$-0.03$ & 0.07 & 29 & 0.02	& 4	\\
45 & 4500 & 1.80 & 1.85 & 5.0 &$-0.05$ & 0.06 & 22 & 0.03	& 4	\\
72 & 4450 & 1.80 & 1.75 & 5.7 &$-0.04$ & 0.09 & 25 & 0.09	& 5	\\
\hline		
	\end{tabular}
\end{table*}

\begin{table*}
\begin{minipage}{180mm}
   \centering
	\caption{Determined abundances and isotopic ratios for programme stars}
	\label{table:4}
	\begin{tabular}{lcccccc}
		\hline\hline
Element	& NGC\,4609	& NGC\,5316	& NGC\,5316	& NGC\,5316	& NGC\,5316	& NGC\,5316\\
		& 43			&	31		&	35		&	45		&	72		& Average*\\
	\hline											
{[C/Fe]}	& $-0.23 \pm 0.01$ (2) & $-0.26 \pm 0.02$ (2) & $-0.10 \pm 0.02$ (2)	& $-0.08 \pm 0.03$ (2) & $-0.20 \pm 0.04$ (2) &$-0.16 \pm 0.07$\\
{[N/Fe]}	& $0.63 \pm 0.02$ (6) &	$0.60 \pm 0.05$	(6)	& $0.82 \pm 0.07$ (6) & $0.50 \pm 0.05$ (6) &	$0.72 \pm 0.08$ (6)	&$0.66 \pm 0.12$\\
{[O/Fe]}	&	$0.11$	(1) 	&	$-0.11$ (1)	&	$0.20$ (1)	&	$0.07$ (1)	&	$0.05$ (1)	&$0.11 \pm 0.11$\\
C/N		&	0.55	&	0.54	&	0.48	&	1.05	&	0.48 &$0.64 \pm 0.24$\\
$^{12}{\rm C}/^{13}{\rm C}$		&	23	&	22	&	23	&	24	&	24	& $23 \pm 1$\\
\noalign{\smallskip}
$\rm [Na/Fe]_{LTE}$	&	$0.42 \pm 0.08$ (4)	&	$0.39 \pm 0.06$ (3)	&	$0.35 \pm 0.03$ (3)	&	$0.33 \pm 0.08$ (3)	&	$0.45 \pm 0.03$ (3)	&$0.38 \pm 0.05$\\
$\rm [Na/Fe]_{NLTE}$	&	$0.33 \pm 0.05$ (3)	&	$0.27 \pm 0.06$	(3)	&	$0.26 \pm 0.05$ (2)	&	$0.22 \pm 0.07$ (3)	&	$0.33 \pm 0.05$ (3)	& $0.27 \pm 0.04$\\
{[Mg/Fe]}	&	$-0.06 \pm 0.01$ (2)	&	$-0.09 \pm 0.05$ (3)	&	$-0.05 \pm 0.06$ (3)	&	$-0.01 \pm 0.11$ (2)	&	$0.04 \pm 0.09$ (3)	& $-0.03 \pm 0.05$\\
{[Al/Fe]}	&	$0.15 \pm 0.01$ (2)	&	$0.11 \pm 0.08$ (2)	&	$0.16 \pm 0.05$ (2)	&	$0.12 \pm 0.01$ (2)	&	$0.12 \pm 0.05$ (2)	& $0.13 \pm 0.02$\\
{[Si/Fe]}	&	$0.13 \pm 0.07$ (10)	&	$0.06 \pm 0.03$ (8)	&	$0.12 \pm 0.04$ (7)	&	$0.13 \pm 0.05$ (7)	&	$0.20 \pm 0.02$ (7)	& $0.13 \pm 0.05$\\
{[Ca/Fe]}	&	$0.07 \pm 0.08$ (7)	&	$0.01 \pm 0.04$ (3)	&	$0.11 \pm 0.05$ (6)	&	$0.10 \pm 0.08$ (5)	&	$0.13 \pm 0.02$ (5) & $0.09 \pm 0.05$	\\
{[Sc/Fe]}	&	$0.01 \pm 0.04$ (5)	&	$0.07 \pm 0.09$ (3)	&	$-0.05 \pm 0.02$ (3)	&	$-0.05 \pm 0.01$ (3)	&	$0.06 \pm 0.03$ (4) & $0.01 \pm 0.06$	\\
{[Ti\,{\sc i}/Fe]}	&	$0.02 \pm 0.07$ (17)	&	$0.02 \pm 0.07$ (12)	&	$-0.02 \pm 0.03$ (10)	&	$0.06 \pm 0.02$ (12)	&	$0.07 \pm 0.06$ (13) & $0.03 \pm 0.04$	\\
{[Ti\,{\sc ii}/Fe]}	&	$0.02 \pm 0.02$ (2)	&	$0.05 \pm 0.02$ (2)	&	$0.03 \pm 0.08$ (2)	&	$0.02 \pm 0.03$ (2)	&	$0.04 \pm 0.01$ (2)&$0.04 \pm 0.01$	\\
{[Cr/Fe]} 	&	$0.06 \pm 0.06$ (10)	&	$0.05 \pm 0.06$ (5)	&	$-0.02 \pm 0.03$ (5)	&	$0.04 \pm 0.09$ (6)	&	$0.09 \pm 0.07$ (5)& $0.04 \pm 0.04$	\\
{[Mn/Fe]}	&	$0.15 \pm 0.06$ (3)	&	$0.12 \pm 0.02$ (4)	&	$0.09 \pm 0.01$ (3)	&	$0.25 \pm 0.04$ (3)	&	$0.18 \pm 0.09$ (3) & $0.16 \pm 0.06$	\\
{[Co/Fe]}	&	$0.13 \pm 0.08$ (4)	&	$-0.02 \pm 0.06$ (6)	&	$0.16 \pm 0.06$ (5)	&	$0.26 \pm 0.05$ (5)	&	$0.06 \pm 0.09$ (2)	& $0.12 \pm 0.11$\\
{[Ni/Fe]}	&	$-0.02 \pm 0.04$ (15)	&	$-0.03 \pm 0.09$ (14)	&	$-0.05 \pm 0.10$ (10)	&	$0.00 \pm 0.10$ (13)	&	$0.07 \pm 0.10$ (18)& $0.00 \pm 0.05$	\\
\noalign{\smallskip}
{[Y/Fe]}	&	$-0.01 \pm 0.06$ (5)	&	$0.22 \pm 0.12$ (4)	&	$0.15 \pm 0.11$ (4)	&	$0.23 \pm 0.09$ (3)	&	$0.12 \pm 0.09$ (3)	& $0.17 \pm 0.05$\\
{[Zr/Fe]}	&	$0.04 \pm 0.08$ (2)	&	$0.10$ (1)		&	$0.26 \pm 0.06$ (2)	&	$0.27 \pm 0.09$ (2)	&	$0.18 \pm 0.09$ (2) & $0.24 \pm 0.04$	\\
{[Ba/Fe]}	&	$-0.09 \pm 0.21$ (2)	&	$0.59 \pm 0.06$ (2)&	$0.30 \pm 0.13$ (2)	&	$0.19 \pm 0.13$ (2)	&	$0.22 \pm 0.17$ (2) & $0.24 \pm 0.05$	\\
{[La/Fe]}	&	$0.05 \pm 0.05$ (3)	&	$0.13 \pm 0.05$ (3)	&	$0.32 \pm 0.10$ (3)	&	$0.37 \pm 0.06$ (3)	&	$0.29 \pm 0.10$ (3)	& $0.33 \pm 0.03$\\
{[Ce/Fe]}	&	$0.03 \pm 0.05$ (3)	&	$0.04 \pm 0.11$ (3)	&	$0.23 \pm 0.13$ (4)	&	$0.20 \pm 0.09$ (4)	&	$0.14 \pm 0.08$ (4)	& $0.19 \pm 0.04$\\
{[Pr/Fe]}	&	$0.07 \pm 0.06$ (2)	&	$0.11 \pm 0.04$ (2)	&	$0.30 \pm 0.07$ (2)	&	$0.31 \pm 0.06$ (2)	&	$0.28 \pm 0.06$ (2) & $0.30 \pm 0.01$	\\
{[Nd/Fe]}	&	$0.10 \pm 0.01$ (4)	&	$0.16 \pm 0.08$ (3)	&	$0.31 \pm 0.11$ (3)	&	$0.30 \pm 0.06$ (3)	&	$0.34 \pm 0.04$ (3) & $0.32 \pm 0.02$	\\
{[Eu/Fe]}	&	$-0.01$ (1)	&	$0.02$ (1) & $0.22$ (1)		&	$0.22$ (1)		&	$0.20$ (1)	& $0.21 \pm 0.01$	\\
\hline
\end{tabular}
\end{minipage}\\
$^\ast$ Star NGC\,5316\,31 was not included in the calculation of average abundance for neutron capture elements.
\end{table*}

The determined atmospheric parameters for the target stars are presented in Table~\ref{table:3}. We determined a metallicity of ${\rm [Fe/H]} = 0.16 \pm 0.08$ for NGC\,4609 which is slightly larger than the only previous photometric determination by \citet{Claria89} who found $\rm [Fe/H]=0.05 \pm 0.13$. As it was mentioned before, NGC\,5316 also had only photometric metallicity determinations ranging from ${\rm [Fe/H]}=0.19$ to ${\rm [Fe/H]}=-0.02$, and our result of ${\rm [Fe/H]}=-0.02 \pm 0.05$ agrees quite well with those studies.

The newly determined metallicities of the clusters and the ages from literature were used to compute the PARSEC isochrones \citep{Bressan12} and to determine the TO masses for the target clusters. We derived a TO mass of $\sim 5.6~M_{\odot}$ with the adopted age of 80~Myr (determined by \citealt{Kharchenko05}) for NGC\,4609, and $\sim 5~M_{\odot}$ with the age of 100~Myr (determined by \citealt{Carraro12}) for NGC\,5316. 

We also have calculated a rough estimate of the Galactocentric distance for NGC\,4609, as no value could be found in previous studies. We took the galactic coordinates, the distance from the Sun derived by \citet{Kharchenko13} ($d=1.3$~kpc), and the Galactocentric distance of 7.98 kpc for the Sun as determined by \citet{Zhu13} from Galactic Cepheids. Our result for the Galactocentric distance of NGC\,4609 is $\sim$7.5~kpc. We also redetermined the galactocentric distance of NGC\,5316 using the distance value from \citet{Kharchenko13}, and we get $ R_{\rm gc} =\sim 7.4$~kpc for NGC\,5316. Our recalculated value is very close to the most recent estimate of 7.6~kpc by \citet{Carraro12}, although they used slightly larger distance values for their determination ($d_{\odot}=1.4$~kpc and $ R_{\rm gc \odot}=8.5$~kpc).

Abundances of chemical elements relative to iron [El/Fe]\footnote{We use the customary spectroscopic notation [X/Y]$\equiv \log_{10}(N_{\rm X}/N_{\rm Y})_{\rm star} -\log_{10}(N_{\rm X}/N_{\rm Y})_\odot$.}, a line-to-line scatter and a number of spectral lines used for the analysis are listed in Table~\ref{table:4}. The ionisation stages of the investigated species can be found in Table~\ref{table:2}. The abundance [Fe/H] stands for 
both Fe\,{\sc i} and Fe\,{\sc ii} abundances since they were required to be the same while determining a value of 
surface gravity in our method of analysis. For oxygen and europium just by one line were used for the analysis thus instead of line-to-line scatteras we may consider the averaged random error which is about $\pm0.06$~dex (see Sect.~\ref{sec:3}). Table~\ref{table:4} also lists the determined C/N and $^{12}$C/$^{13}$C ratios and averaged results for the open cluster NGC\,5316.      

\subsection{Light elements}

\subsubsection{Sodium}

\begin{figure}
	\includegraphics[width=\columnwidth]{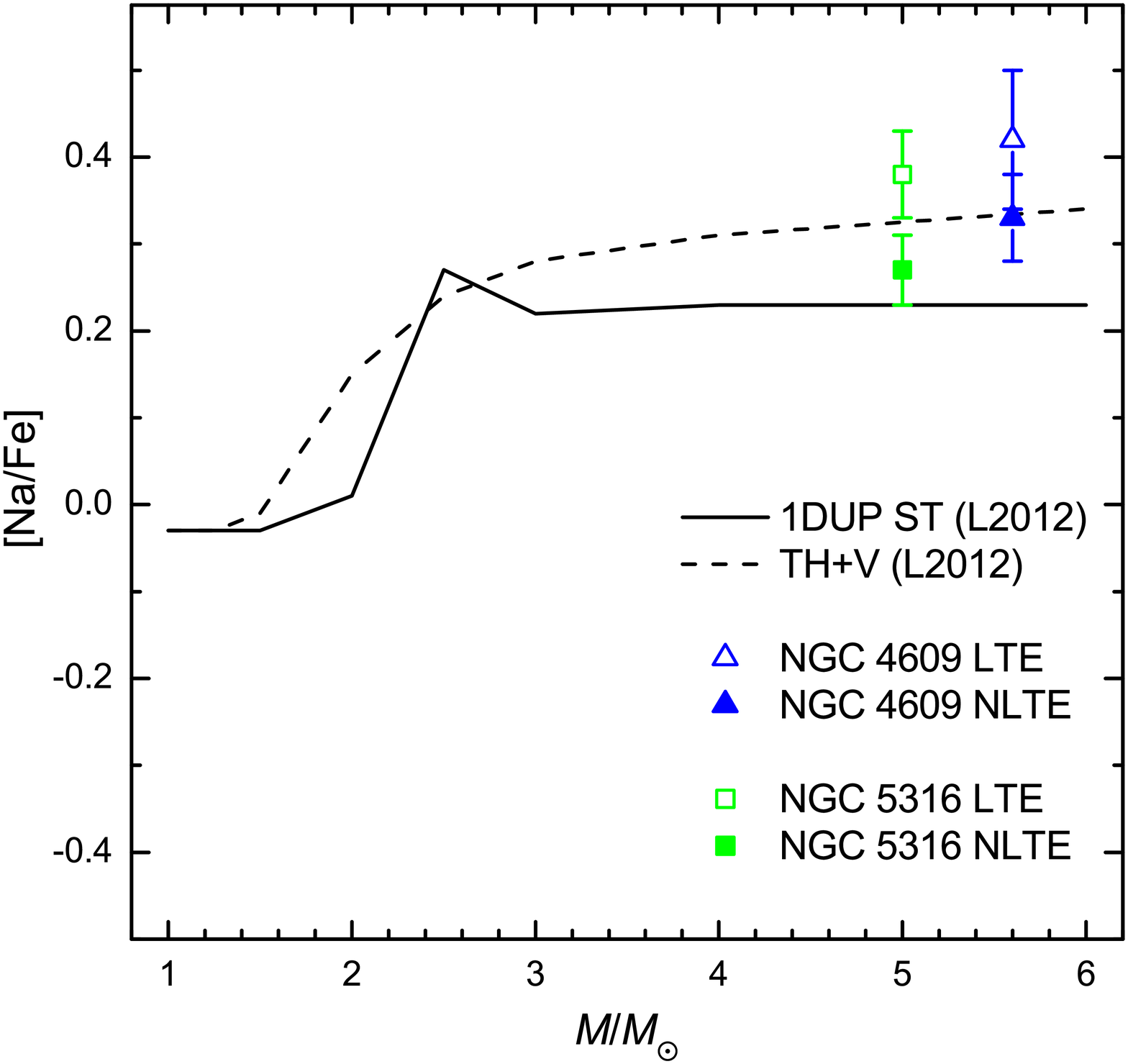}
    \caption{Mean [Na/Fe] values in our sample clusters compared with theoretical extra-mixing models by \citet{Lagarde12}.}
    \label{fig:5}
\end{figure}

Theoretical models which describe chemical mixing in evolved stars \citep{Lagarde12} predict an increase of sodium abundance depending on the turn-off mass. In the recent study by \citet{Smiljanic16} it was shown that this is exactly the case. They plotted seven clusters with [Na/Fe] determinations and made cautious conclusions about the increasing surface sodium abundances with increasing stellar mass. As we can see in Fig.~\ref{fig:5}, our clusters follow the theoretical models as well. Both LTE and NLTE values for the target clusters seem to indicate that the model which includes thermohaline- and rotation-induced mixing \citep{Lagarde12} agrees with the observed values well.

\subsubsection{$\alpha$-elements}

\begin{figure}
	\includegraphics[width=\columnwidth]{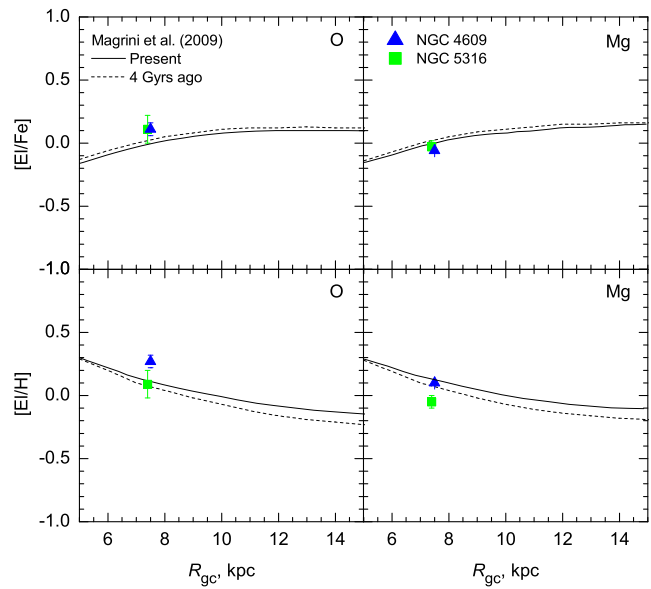}
    \caption{Oxygen and magnesium abundances compared to theoretical models by \citet{Magrini09}.}
    \label{fig:6}
\end{figure}

In Fig.~\ref{fig:6} we compare our two cluster results with theoretical models described by \citet{Magrini09} and provided to us privately. All the abundance ratios in the model are normalized to the time and place of formation of the Sun. Further details can be found in the mentioned paper. Our results agree quite well with the [El/Fe] models. The value for [O/H] is somewhat higher for NGC\,4609 and the value for [Mg/H] is somewhat lower for NGC\,5316, but these deviations are not significant (see e.g. \citealt{Cunha16} who presented results for 29 open clusters). When we calculate the average [$\alpha$/Fe] = \( \frac{1}{4} \) ([Mg/Fe] + [Si/Fe] + [Ca/Fe] + [Ti/Fe]) we get values of 0.04~dex for NGC\,4609 and 0.06~dex for NGC\,5316.  These results are consistent with the thin disk $\alpha$-element abundances.

\subsection{Neutron-capture elements}

The chemical composition of the three programme stars (35, 45, 72) in the cluster NGC\,5316 is rather homogeneous. However, a noticeable difference is seen in the star NGC\,5316\,31 (see Table.~\ref{table:4}). The main s-process element barium is significantly overabundant compared to the other stars and to other s-process element abundances in this star itself.  We rule out errors in the atmospheric parameter determination for this star and a large sensitivity of Ba\,{\sc ii} lines to the microturbulent velocity being a cause of the unusual Ba abundance. The Ba\,{\sc ii} lines in this star are visually stronger in comparison to profiles of other s-process element lines which are also ionised.   

Contrary to barium, other s- and r-process chemical elements (except yttrium), show a slight underabundance compared to other stars of this cluster. However, NGC\,5316\,31 does not fall into a category of barium-rich stars as defined by \citet{Castro16}. \citet{Dias01} give an 86\% probability for this star being a member of NGC\,5316. Its radial velocity determined in our work as well as in studies by \citet{Mermilliod08}, \citet{Frinchaboy08} and others attribute the cluster membership for this star,  thus we conclude that this star is an anomalous one. For these reasons we do not include the neutron-capture chemical elements of this star when calculating the average abundances of elements for the whole cluster as we conclude that this anomaly does not reflect the chemical composition of this cluster. Abundances of other chemical elements in this star agree well with those in other stars of this cluster. 

\begin{figure}
	\includegraphics[width=\columnwidth]{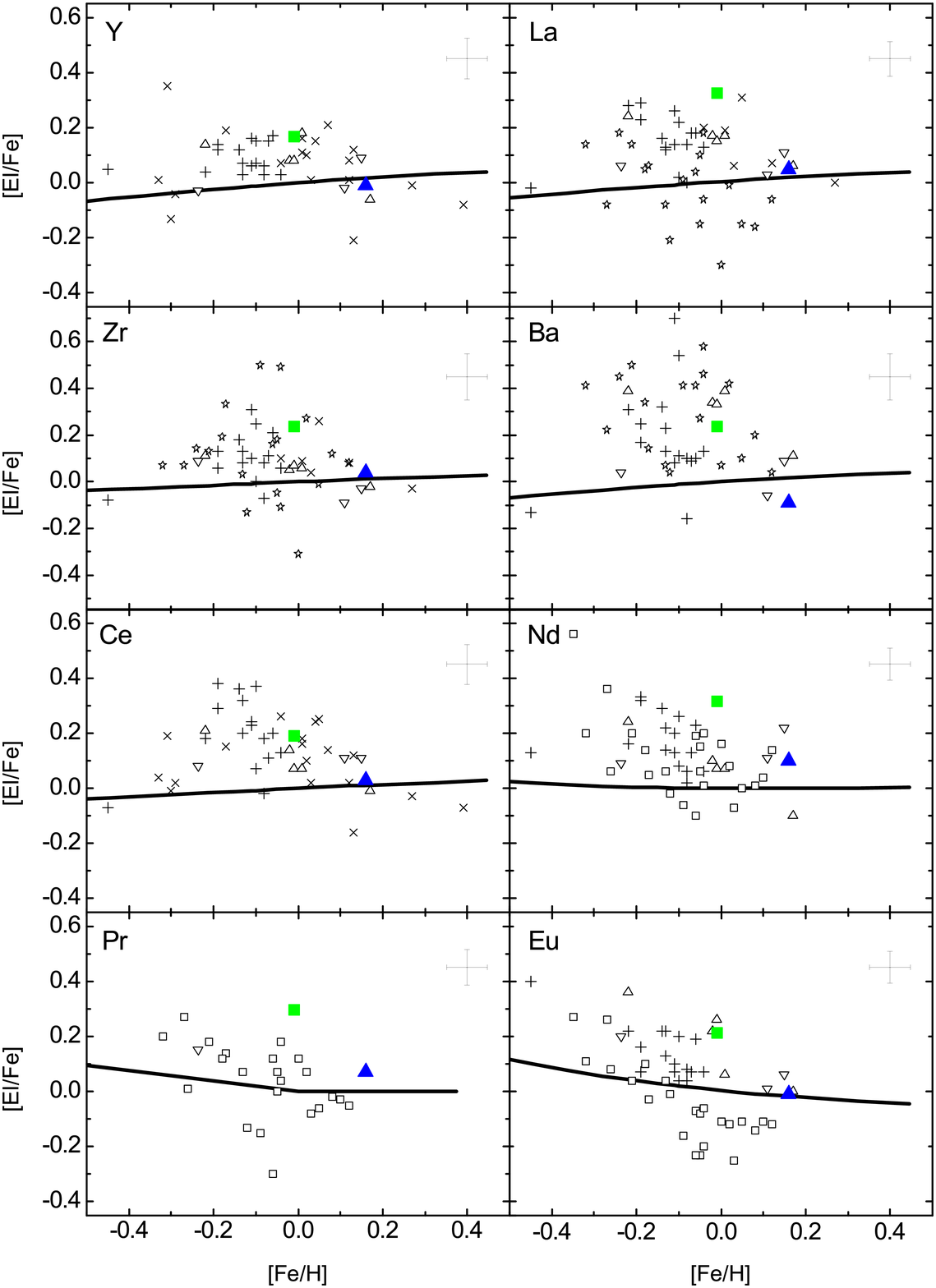}
    \caption{Neutron-capture element abundances of our target clusters compared to previous studies and the models for the thin disk stars by \citet{Pagel97}. Our clusters are indicated as filled symbols: a square for NGC\,5316 and a triangle for NGC\,4609. Plus signs indicate results by \citet{Reddy12, Reddy13, Reddy15}; triangles - results by \citet{Mishenina15}; crosses - results by \citet{Maiorca11}; stars - results by \citet{Jacobson13}; open squares - \citet{Overbeek16} and open reverse triangles indicate results by \citet{Mikolaitis10, Mikolaitis11a, Mikolaitis11b}}
    \label{fig:7}
\end{figure}

For heavy chemical elements which are attributed to the s-process, we derived abundances of Y, Zr, Ba, La, Ce, and Nd. As described in \citet{Travaglio04}, 74\% of yttrium and 67\% of zirconium in the Sun are made by s-process reactions. For Ba, La, Ce, Nd, the s-process reactions contribute by 81\%, 62\%, 77\%, 56\%, respectively \citep{Arlandini99}. 

Fig.~\ref{fig:7} shows the average abundances of neutron-capture chemical elements of our clusters as well as the average cluster results from other studies. 
We summarized the following previous studies: \citet{Maiorca11, Jacobson13, Reddy12, Reddy13, Reddy15, Mishenina15, Mikolaitis10, Mikolaitis11a, Mikolaitis11b, Overbeek16}. We also show a thin disk chemical evolution model developed by \citet{Pagel97}. We do not display results of Eu obtained by \citet{Jacobson13} since they were redetermined by \citet{Overbeek16} from the same spectra. The error-bars presented for every element are median errors computed from scatters around the mean values in all the studies used in plotting this figure.  

NGC\,5316 shows a good match with other studies and fails to concur the thin disk model. This result is the same as found in the majority of other studied open clusters and confirms the idea that s-process chemical elements in clusters are higher than those in field stars (\citealt{Marsakov16} and references therein). As previously mentioned, the enrichment of AGB stars with s-process elements via the ($^{13}{\rm C}(\alpha,n)^{16}{\rm O}$) reaction is most visible in young open clusters (\citealt{Maiorca11, D'Orazi09, Mishenina15} etc.). This could explain the overabundance of these elements in NGC\,5316, which is only 100 Myr old, compared to the thin disc. 

The abundances of neutron-capture elements in NGC\,4609 are lower by about 0.2~dex than in NGC\,5316. Its abundances of s-process elements are almost the same as in the thin disk stars. This difference in abundances of two of our programme clusters is interesting, because both clusters are of a similar age, Galactocentric distance and even are at almost the same place in the Galaxy. However, one has to take into account, that NGC\,4609 has only one identified red giant star, and that is the only star we have analysed in our work.  

We determined abundances of  praseodymium and europium which are attributed to the rapid neutron capture process (r-process) elements produced during supernovae explosions. According to theoretical studies (\citealt{Arlandini99} and others) around 51\% of praseodymium and 91\% of europium are produced during such events.  NGC\,5316 has higher values of Pr and Eu compared to the thin disc. NGC\,4609 agrees better with the thin disk models and shows solar ratios for r-process elements.

\subsection{$^{12}{\rm C}/^{13}{\rm C}$ and C/N ratios}

\begin{figure}
	\includegraphics[width=\columnwidth]{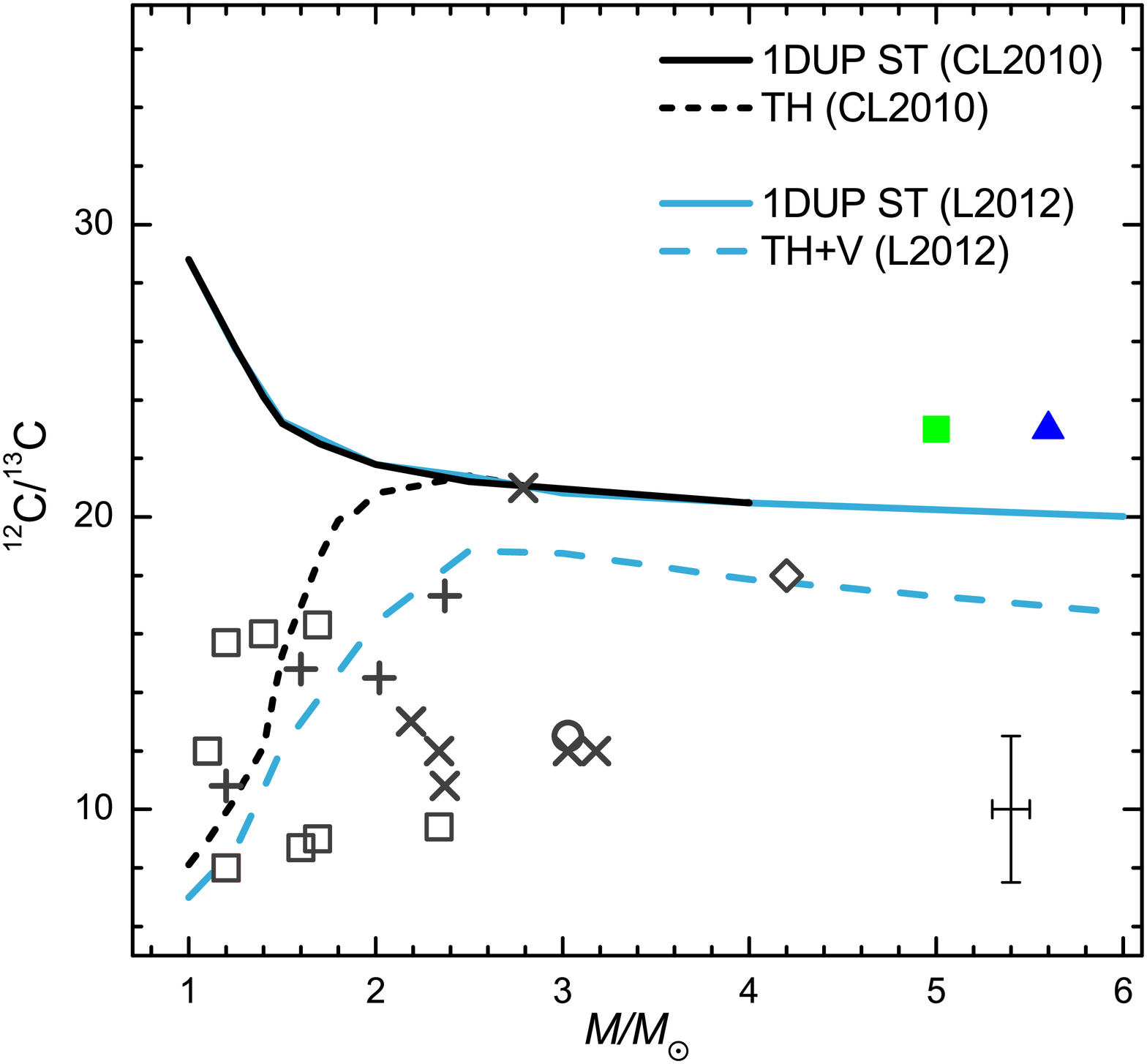}
\caption{The average carbon isotope ratios in clump stars of open clusters as a function of stellar  turn-off mass. Filled square indicates the value for NGC\,5316, and filled triangle -- for NGC\,4609. Open squares indicate previous results by \citet{Tautvaisiene00, Tautvaisiene05, Mikolaitis10, Mikolaitis11a, Mikolaitis11b, Mikolaitis12, Drazdauskas16}. Other symbols include results from \citet{Gilroy89} -- pluses, \citet{Luck94} -- open circles, \citet{Smiljanic09} -- crosses, \citet{Santrich13} -- open diamond. The solid lines (1DUP ST) represent the $^{12}{\rm C}/^{13}{\rm C}$ ratios predicted for stars at the first dredge-up with standard stellar evolutionary models of solar metallicity by \citet{Charbonnel10} (black solid line) and \citet{Lagarde12} (blue solid line). The short-dashed line (TH) shows the prediction when just thermohaline extra-mixing is introduced (\citealt{Charbonnel10}), and the long-dashed line (TH+V) is for the model that includes both the thermohaline and rotation induced mixing (\citealt{Lagarde12}). A typical error bar is indicated (\citealt{Charbonnel10, Smiljanic09, Gilroy89}).} 
    \label{fig:8}
\end{figure}

\begin{figure}
	\includegraphics[width=\columnwidth]{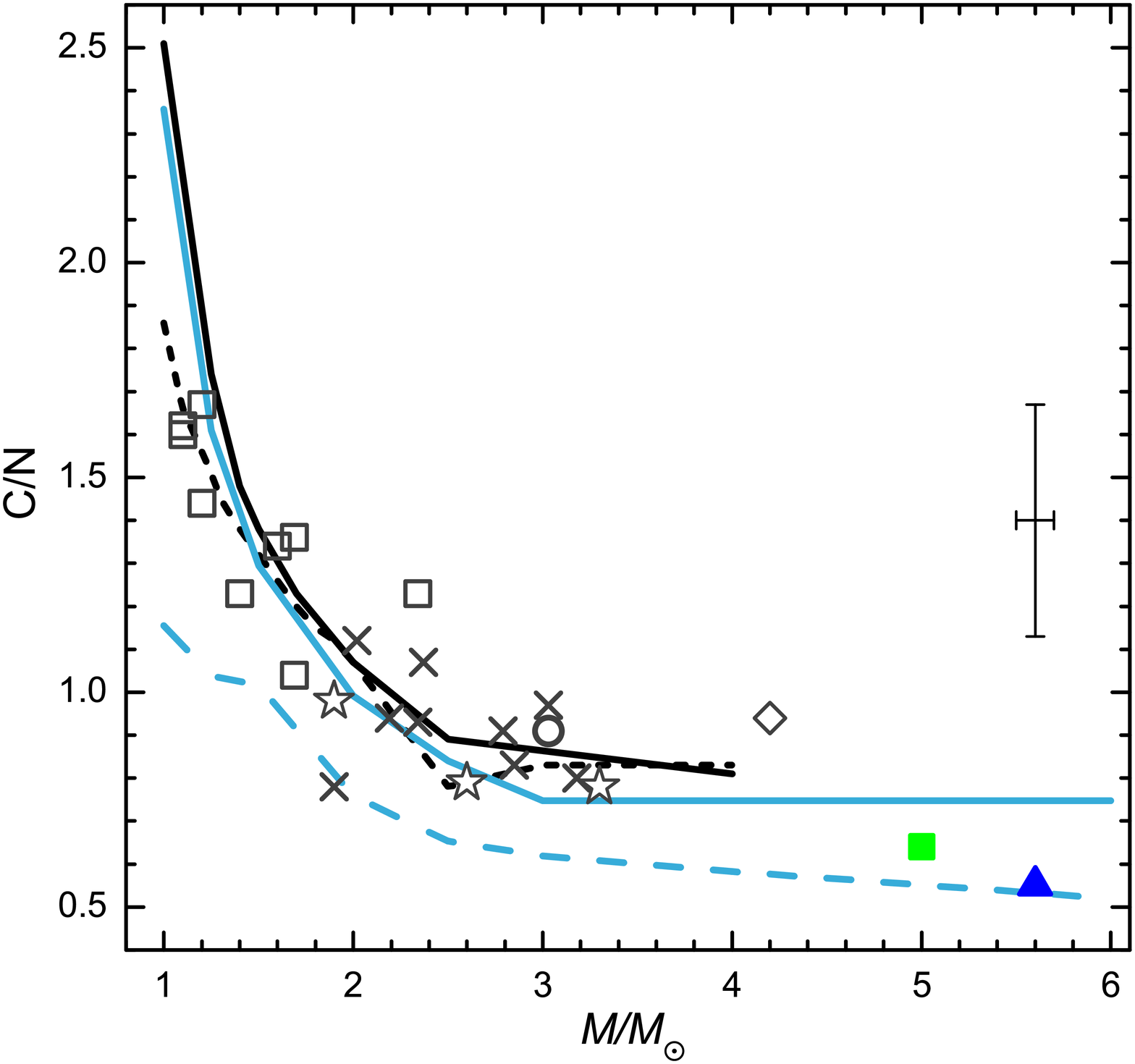}
    \caption{The average carbon-to-nitrogen ratios in clump stars of open clusters as a function of stellar  turn-off mass. In addition to symbols in Fig.~\ref{fig:8} here we include the results from \citet{Tautvaisiene15} as open stars.}
    \label{fig:9}
\end{figure}

As already mentioned in the introduction of this paper, abundances of carbon and nitrogen together with C/N and especially $^{12}{\rm C}/^{13}{\rm C}$ ratios are important tools when trying to understand stellar evolution. All the accumulated chemical changes that happen in stars during the RGB phase are reflected in the atmospheres of He-core  burning stars. Studying these changes can provide valuable insight into stellar evolution. Now it is clear, that the first dredge-up models alone cannot account for all the observed abundance changes. \citet{Eggleton08} called this extra-mixing a $\delta \mu$ mixing, and concluded that it is a significant and inevitable process in low mass stars that are ascending the red giant branch for the first time. They could not accurately define the speed of this mixing, but showed that it is relatively fast compared to the nuclear time scale. Their models, at the time, accounted well for the observed decrease in $^{12}{\rm C}/^{13}{\rm C}$ ratios than predicted before.

The recent models by \citeauthor{Charbonnel10} and \citeauthor{Lagarde12} are based on the ideas of \citet{Eggleton06}, \citet{Ulrich72}, \citet{Charbonnel07}, \citet{Kippenhahn80}. In their study \citeauthor{Eggleton06} found a mean molecular weight inversion ($\mu$) in $\rm 1M_{\sun}$ stellar evolution model. It occurred right after the so-called luminosity bump on the RGB, when the H-burning shell encounters the chemically homogeneous part of the envelope. \citet{Ulrich72} predicted, that this {$\mu$}-inversion is the outcome of the $^3{\rm He(}^3{\rm He}, 2p)^4{\rm He}$ reaction. It does not occur earlier because the {$\mu$}-inversion is low and negligible compared to a stabilising {$\mu$}-stratification. \citet{Charbonnel07} used the ideas by \citeauthor{Ulrich72} (which were extended to non perfect gas by \citealt{Kippenhahn80}), computed stellar evolution models introducing a double diffusive instability called thermohaline convection and showed its importance in the evolution of red giants. This mixing process connects the external wing of the hydrogen burning shell to the convective envelope and induces element abundance changes in the atmospheres of evolved stars.

\citet{Lagarde12} further developed these models adding the effects of rotation-induced mixing. Typical initial zero-age main sequence rotation velocities were chosen which depend on stellar mass and are based on observed velocity distributions in young open clusters \citep{Gaige93}. In the models, the convective envelope rotates as a solid body, and the transport coefficients associated with the thermohaline- and rotation-induced mixing were simply added in the diffusion equation. The rotation-induced mixing modifies the internal chemical structure even before the RGB phase, but the results are only visible in later evolutionary stages.

Figs.~\ref{fig:8}--\ref{fig:9} display the comparison between theoretical models and  $^{12}{\rm C}/^{13}{\rm C}$ and $\rm C/N$ ratios of stars in different open clusters as a function of TO mass. The theoretical models include the first dredge-up, thermohaline (TH), and thermohaline and rotation (TH+V) induced mixing computed by \citet{Charbonnel10} and \citet{Lagarde12}. We included the observational results from other studies as well \citep{Gilroy89, Luck94, Tautvaisiene00, Tautvaisiene05, Tautvaisiene15, Mikolaitis10, Mikolaitis11a, Mikolaitis11b, Mikolaitis12, Smiljanic09, Drazdauskas16}.

The turn-off masses for our clusters are 5.6 and 5~$M_{\odot}$ for NGC\,4609 and NGC\,5316, respectively. All stars in our sample are considered as being the He-core burning stars. Compared with the theoretical models we can see that in this mass range extra-mixing does not have a significant effect and our mean $^{12}{\rm C}/^{13}{\rm C}$ values agree with the standard first dredge-up and thermohaline-induced mixing model (\citealt{Charbonnel10})  and are not lowered as much as predicted by the models where thermohaline- and rotation-induced mixing act together (\citealt{Lagarde12}). The result of open cluster NGC\,3114  by \citet{Santrich13} lies below the first dredge-up model. Certainly, more clusters with large TO masses should be observed in order to have a clearer picture. The same is valid in case of C/N ratios (Fig.~\ref{fig:9}).  
C/N ratios have larger uncertainties, our results for both clusters are very close and seem to agree with both the first dredge-up and the thermohaline- and rotation-induced mixing models. The result of NGC\,3114 this time is more close to the first dredge-up model.  

This study further shows that we need more observational data to better understand the mixing processes in evolved stars and to make more precise models.

\section{Conclusions}

Two young open clusters were analysed for the first time using high resolution spectroscopy. The main atmospheric parameters and abundances of 23 chemical elements were determined.  The main conclusions of our chemical composition analysis can be summarized as follows:
\begin{enumerate}
\item
NGC\,4609 has a slightly larger than solar metallicity of ${\rm [Fe/H]} = 0.16 \pm 0.08$, a turn-off mass of about 5.6~$M_{\odot}$ (with the age of 80~Myr), and a Galactocentric distance of about 7.5~kpc.  
\item
NGC\,5316 is of solar metallicity, its ${\rm [Fe/H]} = -0.02 \pm 0.05$, the turn-off mass is about 5~$M_{\odot}$ (with the age of 100~Myr). 

\item 
$^{12}{\rm C}/^{13}{\rm C}$ and C/N ratios and [Na/Fe] agree quite well with the model which takes into account thermohaline- and rotation-induced mixing but within error limits also agree with the standard first dredge-up model.
\item
 Comparison of oxygen, magnesium and other $\alpha$-elements with theoretical models of Galactic chemical evolution revealed that both clusters follow the thin disk $\alpha$-element trends.
\item
The open cluster NGC\,5316 confirms the enrichment s-process element abundances seen in other young open clusters compared to old clusters. The star NGC\,5316\,31 has an exceptionally high barium abundance. NGC\,4609 has solar neutron-capture element-to-iron ratios.
\end{enumerate}

There are few clusters of intermediate turn-off masses for which a detailed chemical composition is studied, therefore  NGC\,4609 and NGC\,5316 add useful data to this small sample.
Further spectroscopic analyses of other young clusters will be welcomed in order to investigate the abundance pattern in evolved intermediate mass stars. 

\section*{Acknowledgements}

This  research  has  made  use  of  the  WEBDA  database (operated  at  the  Department  of  Theoretical  Physics  and  Astrophysics  of  the Masaryk  University,  Brno),  of  SIMBAD  (operated  at  CDS,  Strasbourg),  of VALD \citep{Kupka00}, and of NASA's Astrophysics Data System. Bertrand Plez (University of Montpellier~II) and Guillermo Gonzalez (Washington State University) were particularly generous in providing us with atomic data for CN and $\rm C_{2}$ molecules, respectively. We thank Laura Magrini for sharing with us the Galactic chemical evolution models. AD, GT, YC, and VB  were  partly  supported by the grant from the Research Council of Lithuania (MIP-082/2015). RS acknowledges support by the National Science Center of Poland through the grant 2012/07/B/ST9/04428.




\bibliographystyle{mnras}
\bibliography{References-Drazdauskas} 


\end{document}